\documentclass[11pt]{article}
\usepackage{blois,epsfig}

\def\PLB{{\em Phys. Lett.}  B}

\def\be{\begin{equation}}
\def\ee{\end{equation}}
\def\bea{\begin{eqnarray}}
\def\eea{\end{eqnarray}}
\def\bec{\begin{center}}
\def\enc{\end{center}}

\def\mrm{\mathrm}
\def\unit{\, \mrm}
\def\lunits{\unit cm^{-2}\unit s^{-1}}

\begin{document}
\vspace*{2cm}
\begin{center}
\Large{\textbf{XIth International Conference on\\ Elastic and Diffractive Scattering\\ Ch\^{a}teau de Blois, France, May 15 - 20, 2005}}
\end{center}

\vspace*{2cm}
\title{Elastic Cross-Section and Luminosity Measurement in ATLAS at LHC}

\author{ I. Efthymiopoulos \\ (for the ATLAS Collaboration)}

\address{CERN, AB Department\\ 1211 Geneve 23, Switzerland}

\maketitle\abstracts{
Recently the ATLAS experiment was complemented with a set 
of ultra-small-angle detectors located in ``Roman Pot'' inserts at 240m on
either side of the interaction point, aiming at the  absolute determination
of the LHC luminosity by measuring the elastic scattering rate at the 
Coulomb Nuclear Interference region. Details of the proposed measurement the 
detector construction and the expected performance as well as the challenges 
involved are discussed here.} 

\section{Introduction}

In March 2004 the ATLAS/LHC collaboration submitted a letter of intent~\cite{loi}
proposing to complement the experiment with ultra-small-angle detectors,
in order to measure elastically scattered protons for the primary purpose of absolute
determination of the LHC luminosity at the ATLAS interaction point(IP). 
The detectors will be housed in ``Roman Pot'' (RP) inserts in the beam pipe
located at $240\unit{m}$ on either side of the IP  and 
will be able to detect the scattered protons at small enough (mm) distances
away from the circulating beam, aiming to reach with the help of special
optics, the theoretically well-calculable Coulomb scattering regime. To
complete the picture, a second detector called LUCID, 
based on cylindrical Cerenkov counters is proposed to be installed around the 
beam pipe close to the interaction point. The LUCID detector is a luminometer, 
which will be calibrated using the RP detectors and would allow to transfer the
luminosity measurement from the low ($10^{27}\lunits$) values used during the
special luminosity runs, to the high ($10^{34}\lunits$) values during the
main data-taking of ATLAS.

Measuring the small-angle elastic scattering at 
the Coulomb region is a very attractive and at the same time very a
challenging  experiment. Other possibilities to accurate estimate and monitor
the LHC luminosity with the ATLAS detector will be pursued as well. For
example, measuring the $W\rightarrow\ell\nu_\ell$ and 
$Z\rightarrow\ell\ell$ rates or the di-muon pair production rate in
double-photon exchange, the relative LHC luminosity could be determined very
accurately due to the large statistics available. However it may be difficult
to reach an absolute error close to a percent, due to the theoretical
uncertainties in the cross-sections and branching ratios
involved. Using the machine parameters and optics to estimate the luminosity
is another possibility that will be investigated, however the expected error
is estimated to be around 10\%, dominated by the knowledge of the beam sizes
around the interaction region.

\begin{figure}
\begin{center}
\begin{minipage}[t]{0.49\textwidth}
  \epsfig{figure=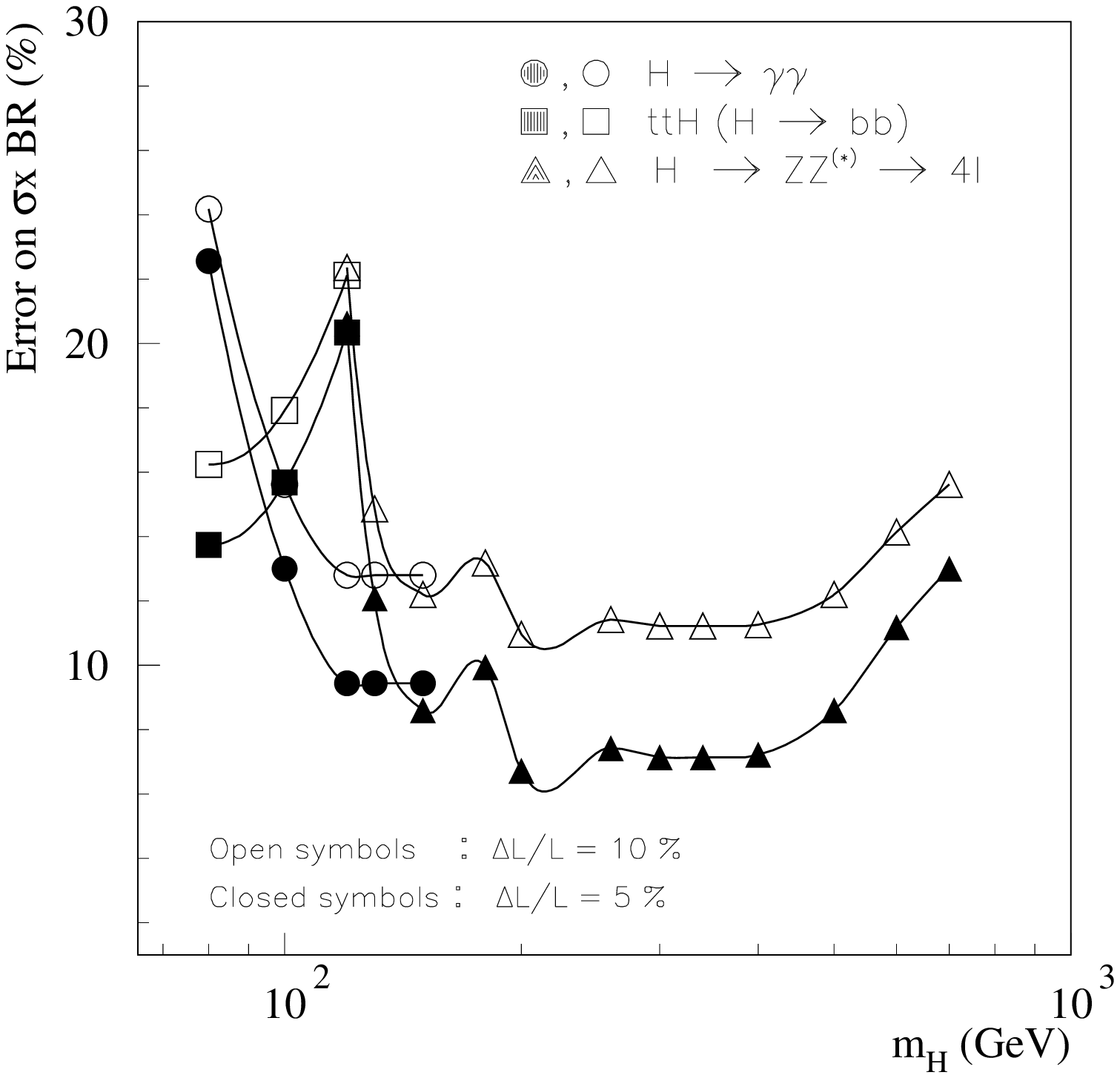,width=0.90\textwidth,height=3.5cm} 
  \caption{\label{fig:hrate} Relative precision on the measurement of the 
      Higgs-boson rate for
      various channels as a function of $m_H$ assuming an integrated luminosity
      of $100\unit{fb^{-1}}$ and for two different luminosity errors: 10\% and 5\%.}
\end{minipage}
\hfill
\begin{minipage}[t]{0.49\textwidth}
  \epsfig{figure=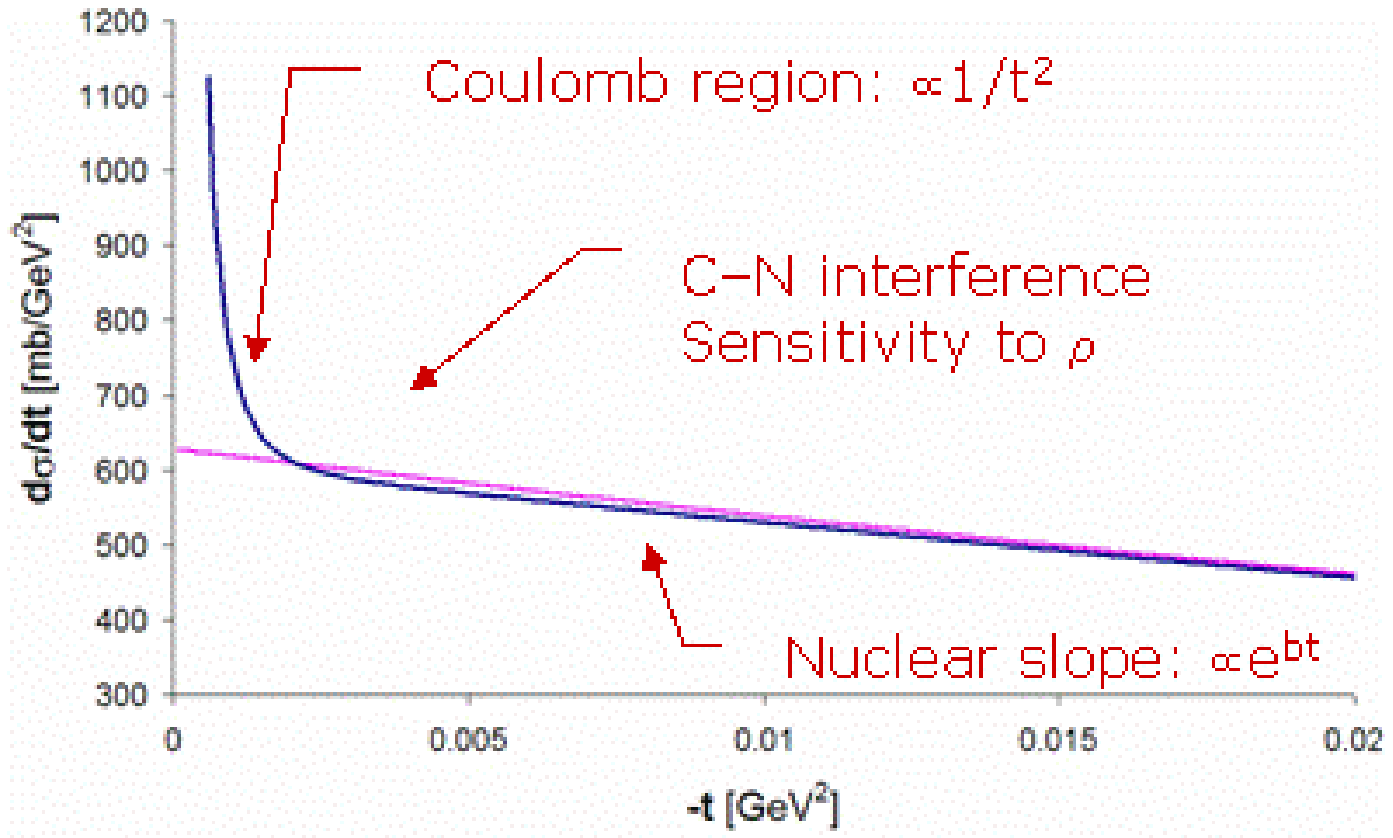,width=0.90\textwidth,height=3.5cm} 
  \caption{\label{fig:xsection} Graphical representation of the total cross-section around the CNI region.}  
\end{minipage}
\end{center}
\end{figure}

An accurate knowledge of the  luminosity would be very desirable for a number of
physics studies in ATLAS~\cite{tdr}. As an example in the measurement of the 
 Higgs-boson rate as shown in Fig.~\ref{fig:hrate}, or in the measurement of cross-sections
for processes like $t\bar{t}$ production, which, if could be very precisely measured and 
found different from the SM predictions, could be the sign of interference with new physics!

Traditionally, for hadron colliders, the luminosity as well as the total cross-section 
has been measured using the so-called ``luminosity independent'' method. 
In this approach one needs to measure both the total interaction ($R_{tot}$) and the 
elastic $dR_{el}/dt$ rates in the forward region. This method was successfully used in previous
experiments and is proposed by TOTEM for
LHC~\cite{totem}. Unfortunately the ATLAS detector has limited coverage in the
forward region to allow a reasonable measurement of the total
inelastic rate. Thus, the only possibility to measure locally 
(i.e. at ATLAS) the absolute luminosity is to go into even smaller angles and 
reach the Coulomb Nuclear Interference (CNI) region. 

\subsection{Elastic Scattering at the Coulomb Nuclear Interference Region}\label{subsec:cni}

For very small momentum transfer values $-t$, $-t=(p\theta)^2$, the cross-section becomes
sensitive to the well-known electromagnetic amplitude (see Fig~\ref{fig:xsection}) as expressed by the formula: 
\begin{equation}
\left.\frac{dN}{dt}\right|_{t\simeq0} = L\pi\left(f_C+f_N\right)^2\simeq
L\pi\left(-\frac{2\alpha_{EM}}{|t|}+\frac{\sigma_{tot}}{4\pi}(i+\rho)e^{-b|t|/2}\right)^2
\label{eq:cni}
\end{equation}
Thus, provided that we are able to probe the CNI region (i.e. the region of
$-t$ values where the strong and the electromagnetic amplitudes become equal)
and with the help of the optical theorem, a simple measurement of the 
elastic rate would allow the determination of the luminosity without
measuring the inelastic rate. Moreover, since in practice we will fit the
data to Eq~\ref{eq:cni}, the other important parameters such as the
$\rho$-parameter, the total cross-section ($\sigma_{tot}$), and the nuclear
slope $b$ will be determined as well. This very powerful technique has been
explored previously in the UA4 experiment~\cite{ua4} at the $Sp\bar{p}S$ collider with
very good results. Our aim is to determine the luminosity within a
2\% error and give a competitive measurement on the other parameters.

\section{The Experimental Setup}

The CNI region at LHC of $7\unit{TeV}$ corresponds to
$|t|=6.5\times10^{-4}\unit{GeV^2}$, or for scattering angle
$\theta_{min}\leq3.5\unit{\mu rad}$. This is a quite challenging angle to
measure, about 40 times smaller from that of UA4 at the SPS. Nevertheless, we
believe it can be done using special optics and running conditions of the LHC
machine, accompanied with very performant edge-less detectors and RP system.

\subsection{Optics and Beam Parameters}\label{subsec:bopt}

The typical optics for an elastic scattering experiment is the so-called
``parallel to point'' optics from the interaction region to the detector
location. In such an optics, the betatron oscillation between the
interaction point of the elastic collision and the detector position has a 90
degree phase difference in the measuring vertical plane
%~\footnote{Since the
%  LHC machine has two rings very close to each other, it is technically
%  difficult to install RP detectors on both sides of the beam pipe at the
%  horizontal plane.} 
such that all particles scattered at the same angle are focused at the same
locus at the detector, independent on their vertex position. In such a case, the measured
coordinate $y_{det}$ at the detector is related to the scattering angle
according to $y_{det}=\theta^* L_{eff}$ while for the momentum transfer $-t$ we obtain:
\begin{equation} \label{eq:tval}
-t = 
%\left( p_{beam}\theta^*\right)^2 = 
\left( p_{beam}\frac{y_{det}}{L_{eff}}\right)^2=
\left( p_{beam}\frac{y_{det}}{\sqrt{\beta\beta^*}}\right)^2
\end{equation}
Expressing the minimum approach $y_{min}$ in terms of
the beam size at the detector 
location ($y_{min}=n_\sigma \sigma_y$) we obtain:
\begin{equation}\label{eq:tmin}
-t_{min}=\left( p_{beam}\frac{n_\sigma\sigma_y}{\sqrt{\beta\beta^*}}\right)^2 =
p^2_{beam}\left( \frac{n_\sigma\sqrt{\beta\epsilon_N/\gamma}}{\sqrt{\beta\beta^*}}\right)^2 =
p^2_{beam} n^2_\sigma \frac{\left(\epsilon_N/\gamma\right)}{\beta^*}
\end{equation}

From Eq.~\ref{eq:tval} and~\ref{eq:tmin} the important
parameters of the experiment in order to reach the lowest possible $t_{min}$
value can be extracted: large $L_{eff}$ which means the detectors must be far away from IP,
large $\beta^*$ which means special optics, small emittance $\epsilon_N$ and
small $n_\sigma$ which means special running conditions and good cleaning
efficiency in the machine collimation system in order to keep 
the beam halo and background rate to affordable levels and allow going close
to the beam.  

\begin{figure}
\begin{center}
\epsfig{figure=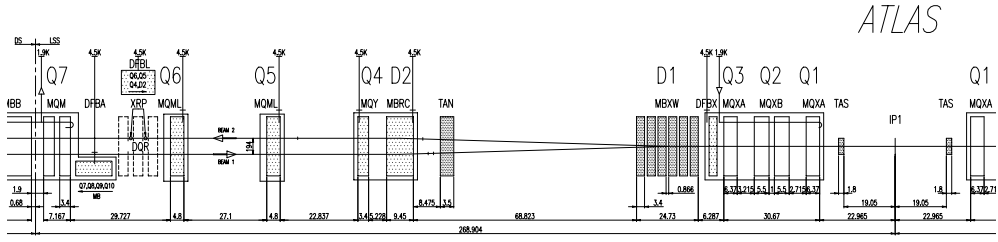,bbllx=2,bblly=348,bburx=588,bbury=518,clip=,height=1.5in} 
\end{center}
\caption{\label{fig:rplayout} Schematic layout of the LHC LSS1 near the ATLAS
  interaction point showing the location for the RP units in the region
  between the Q6 and Q7 quadrupoles.}
\end{figure}

\begin{figure}
  \begin{minipage}[c]{0.60\textwidth}
    \epsfig{figure=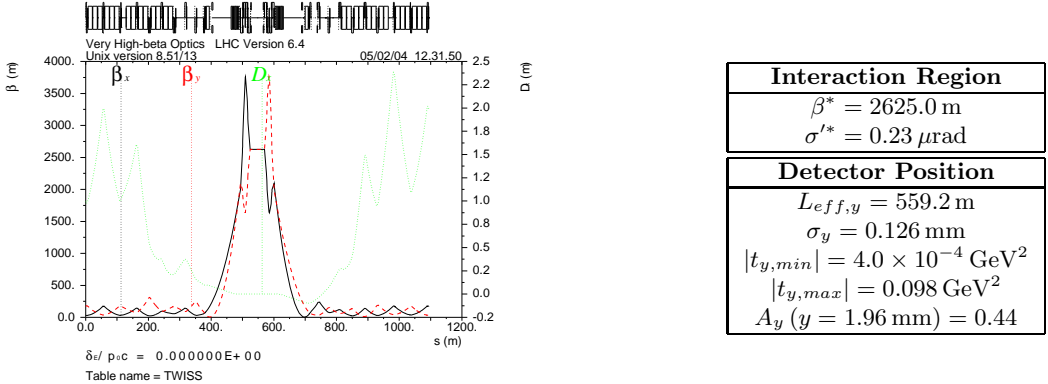,clip=,width=5.5cm,angle=-90} 
  \end{minipage}
  \hfill
  \begin{minipage}[c]{0.40\textwidth}
{\footnotesize
	\begin{tabular}{|c|}
	  \hline
	  {\bf Interaction Region}\\ \hline
	      $\beta^*= 2625.0\unit{m}$ \\
	      $\sigma^{\prime *} = 0.23\unit{\mu rad}$ \\ \hline \hline
	  {\bf Detector Position} \\ \hline
	  $L_{eff,y} = 559.2\unit{m}$ \\
	  $\sigma_{y} = 0.126\unit{mm}$ \\ 
	  $|t_{y,min}| = 4.0\times10^{-4}\unit{GeV^2}$ \\ 
	  $|t_{y,max}|=0.098\unit{GeV^2}$ \\
	  $A_{y}\,(y=1.96\unit{mm}) = 0.44$ \\ \hline
	\end{tabular} 
}
  \end{minipage}
\caption{\label{fig:optics}  The optics diagram around IP1 for the very
  high-beta, $\beta^*=2625\unit{m}$, optics solution.  The main beam parameters 
are also shown at the ATLAS IP and at the RP detector location at $240\unit{m}$ distance. 
The beam parameters were calculated using a normalized transverse emittance of $1\unit{\mu m\, rad}$.} 
\end{figure}

For our experiment, a location in the LHC machine between the Q6 and Q7
quadrupoles at 240 meters from the ATLAS/IP1 was found, where
a station of two RP units separated by 3.5~m could be installed as shown in 
Fig.~\ref{fig:rplayout}.
In parallel, a ``high-beta'' optics solution was found that
satisfies the experimental requirements using the existing LHC magnetic
elements within the specifications~\cite{optics}. The optics diagram and the main beam 
parameters are summarized in Fig.~\ref{fig:optics}. To switch from the standard low-beta optics ($\beta^*=0.5\unit{m,}$) to that one, no hardware modifications are required, apart
from inverting the polarity for the Q4 quadrupole that could be done remotely.  

Although the optics would allow reaching the CNI region, the difficulties involved should 
not be underestimated. The
operation scenario envisaged is to have dedicated low intensity runs with only
43 bunches in the machine of $10^{10}$ protons each, in order to achieve
the required transverse emittance of $1.0\times10^{-6}\unit{\mu m\,rad}$ and
cleaning efficiency of the collimators. It should be noted that this emittance is the one
foreseen for the pilot beam~\cite{lhctdr}. In recent SPS machine tests, values
close to the ones required here were obtained for an LHC structured (25~ns
spacing) beam. Of course it remains challenging to preserve the emittance
values after injection at LHC, but we hope that after some experience is
gained could be feasible. Under these running conditions, the RP detectors could 
approach the beam as close as $10\sigma$, staying in the shadow of the machine collimators. 
The accidental halo rate expected is about $6\unit{kHz}$, compared to the expected rate for elastic 
events of $30\unit{Hz}$ during these special low luminosity ($10^{27}\lunits$) runs.

\subsection{Roman Pots and Detectors}\label{subsec:rpdet}

The RP units and the detectors have to satisfy similar stringent
requirements. The most critical concerns the thickness of the pot window and the detectors to be
edge-less in order
to reduce the acceptance loss at the inmost side near the beam. Additional requirements 
for the detectors are for the spacial and angular resolution projecting back to the interaction 
point from the two RP stations to be both better than $30\unit{\mu m}$. 
Radiation hardness is not really an issue since the detectors will be used
only during the low luminosity runs. The RP unit itself is a high precision mechanical system
that should be able to position the two pots (up and down to the beam) with
few $10-15\unit{\mu m}$ precision. Moreover, their design  must respect all the
requirements arising from the LHC machine and due to interferences with the
circulating beam (such as RF interference, vacuum, interlock system,
etc.). For ATLAS, we plan to use the same RP unit design developed by TOTEM. to whom we are very grateful.

\begin{figure}
\begin{center}
\begin{minipage}[t]{0.49\textwidth}
\epsfig{figure=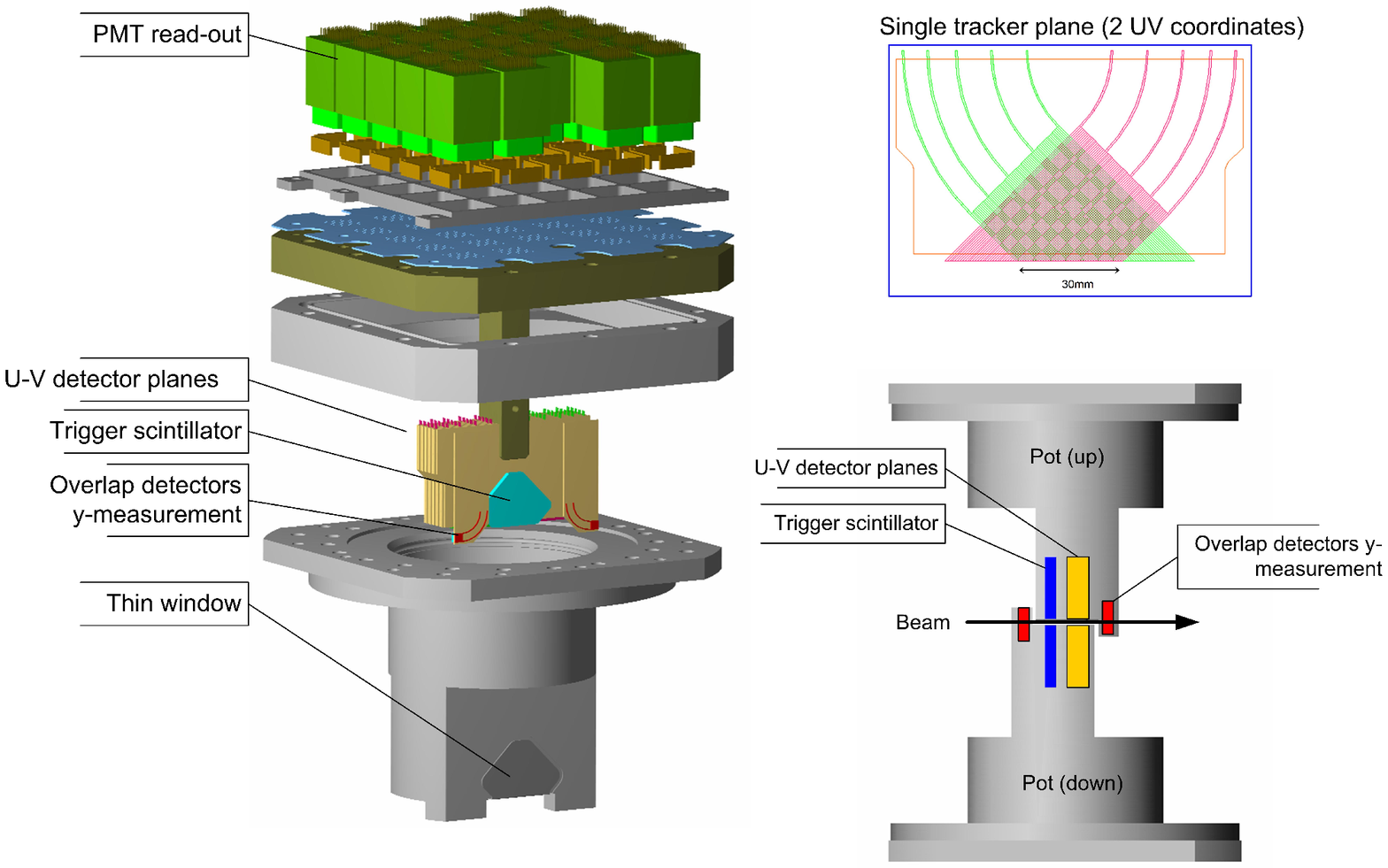,width=0.90\textwidth,height=5.cm} 
\end{minipage}
\hfill
\begin{minipage}[t]{0.49\textwidth}
\epsfig{figure=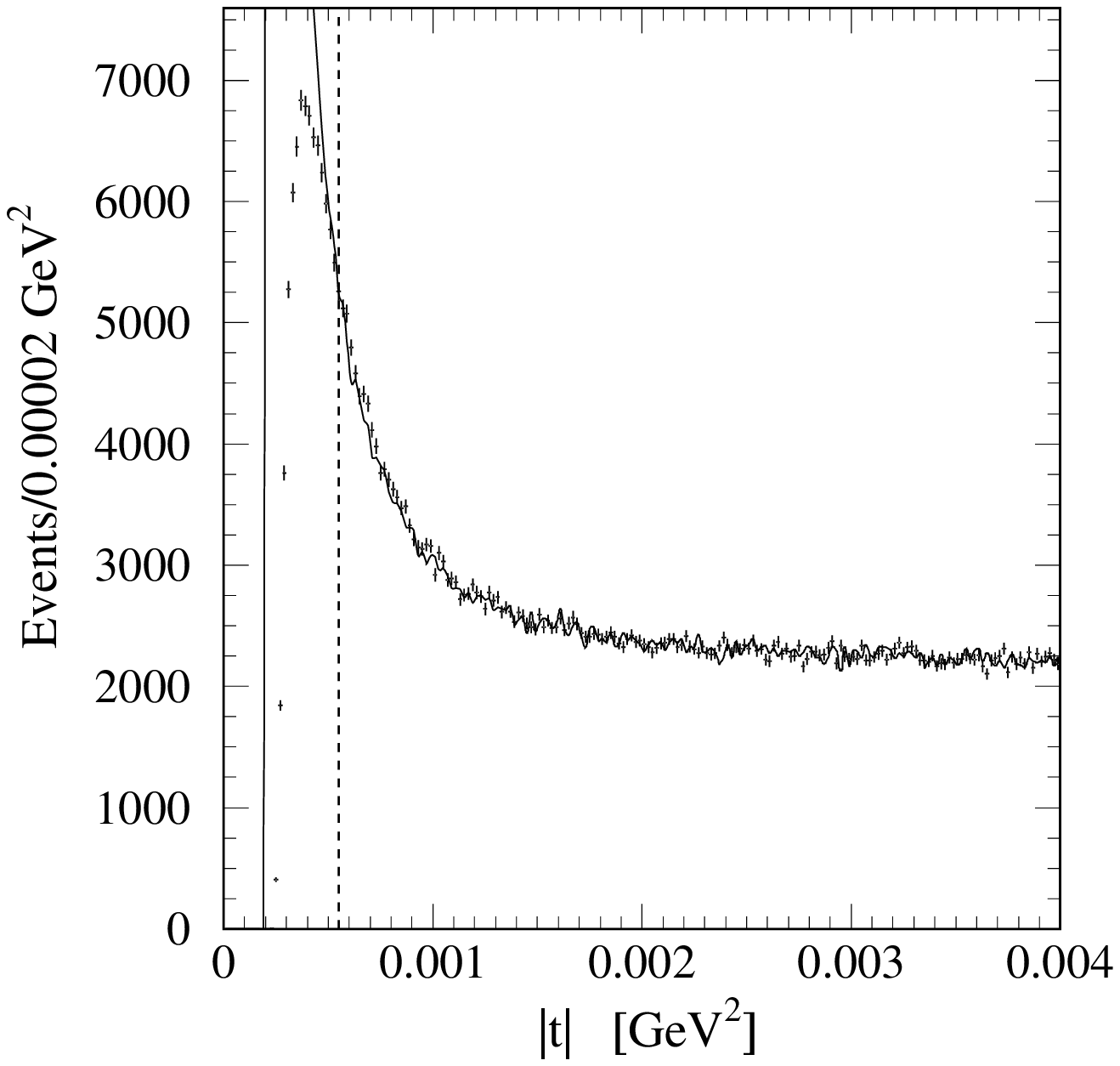,width=0.90\textwidth,height=5.cm}
\end{minipage}
\caption{\label{fig:rpdet} 
Left:Schematic view of the fiber tracker and the pot
  assembly. Right: $dN/dt$ distribution of simulated
  reconstructed events as a function of the reconstructed $-t$ value. A
  fiducial cut of $\pm45$ degree in azimuthal coordinate was applied. The
  solid line indicates the fitted region from $0.00055\leq |t| \leq 0.03\unit{GeV^2}$.}
\end{center}
\end{figure}

Following a review of various options and possibilities, the choice was made
to use scintillating $0.5\unit{mm}$ square fibers bunched to form tracker
planes as detectors read out using multi-anode PMTs. Ten such planes per 
coordinate staggered with a pitch of $50\unit{\mu
  m}$ will form a detector assembly, housed in a single RP as shown in
Fig.~\ref{fig:rpdet}. The
trigger is provided by a large scintillator plane covering the full 
detector area. An elastic event will be identified using an left-up(down) right-down(up) coincidence.
An important aspect in the design is
the two overlap detectors located on each side of the pot in the horizontal
plane which using halo tracks would provide a very precise calibration of the gap between the
two (up and down) pots and therefore of the absolute $-t$ scale. 

From simulations and first laboratory results, about 3 to 5 photoelectrons
are expected per hit, while a resolution of $20\mu m$ with 95\% efficiency was  
achieved. In Fig.~\ref{fig:rpdet} the simulated data on the detector planes are
shown. For each event, the scattering angle from the left and right arm with
respect to IP was reconstructed and combined. As a preliminary result,
using 5 million events, corresponding to about 90 hours of LHC running at
$10^{27}\lunits$  a statistical error of about 2\% in the
luminosity and the other parameters was obtained from a simple fit into the
data. In the final analysis however, the fitting procedure would be more
complicated taking into account the correlation between the parameters as
well as other systematic uncertainties.  

\section{Future plans}

Although our primary goal is to measure the luminosity and the total cross section, 
further studies in order to extend the measurement of the elastic rate to
the maximum $-t$ values are ongoing. Using the very high-beta optics we hope to reach
the CNI region up to $-t$ values of $\sim 0.1\unit{GeV^2}$. Using an intermediate beta
optics and short runs with low intensity, the region $0.1-1.0\unit{GeV^2}$ could be
explored, while going to even higher $-t$ values beyond $1.0\unit{GeV^2}$ high luminosity
runs with low beta optics would be needed, most likely with different
(radiation hard) detectors. Issues like
background estimates, trigger and detector performance should be addressed as
well but the potential of a very interesting physics program is quite
appealing.  

Going even further, at a longer time scale, after sufficient experience in dealing with the difficulties 
associated with working at small distances from the extremely powerful LHC beam 
is gained, it would allow us to propose a  
competitive diffractive physics program using possibly additional detectors 
accessing the full kinematic range of the scattered particle.

\end{document}